\documentclass[acmsmall,screen,nonacm]{acmart}
\usepackage[T1]{fontenc}
\usepackage{graphicx}
\usepackage{subcaption}
\usepackage{amsmath}
\usepackage{booktabs}
\usepackage{tikz}
\usetikzlibrary{shapes.geometric}
\usepackage{listings, listings-rust}
\usepackage[normalem]{ulem}
\usepackage{hyperref}
\usepackage{cleveref}
\usepackage{semantic}
\usepackage{tlatex}
\usepackage{marvosym}

\lstnewenvironment{rustcodenonum}{%
    \lstset{
      language=Rust,
      style=boxed,
      basicstyle=\fontsize{8}{9}\selectfont\ttfamily,
      mathescape=true,
      escapeinside={(*}{*)},
      tabsize=2,
      numbers=none,
    }
}{}
\lstnewenvironment{rustcode}[1][1]{%
    \lstset{
      language=Rust,
      style=boxed,
      basicstyle=\fontsize{8}{9}\selectfont\ttfamily,
      mathescape=true,
      escapeinside={(*}{*)},
      tabsize=2,
      firstnumber=#1,
    }
}{}
\newcommand{\code}[1]{\text{\lstinline[
  language=Rust,
  style=colouredRust,
  basicstyle=\small\ttfamily,
  mathescape=true,
  escapechar=~,
]`#1`}}

\def\always{\square}

\def\eventually{\lozenge}

\makeatletter
\newcommand*\codepoint[1]{\raisebox{1pt}[0.5em][0pt]{\tikz[baseline=(char.base)]{
  \node[
    fill=white,
    shape=circle,
    draw,
    inner sep=1pt,
    minimum width=10pt,
    scale=0.8,
  ] (realchar) {\textnormal{\textbf{\scriptsize\vphantom{WAH1g}#1}}};
  \node[
    inner sep=0pt,
    minimum width=0em,
  ] (char) {\phantom{\texttt{mmm}}};
}}}
\makeatother

\newcommand\codehl[1]{%
  \makebox[0pt][l]{%
    \hspace{-0.5em}%
    \fboxsep0pt%
    \colorbox{red!10}{%
      \hspace{0.5em}%
      \fontsize{8}{9}\selectfont\ttfamily\strut%
      \phantom{\texttt{#1}}%
      \hspace{0.5em}%
    }%
  }}

\newcommand*{\codeltl}[1]{%
  \hspace{-0.25em}%
  \fboxsep0pt%
  \colorbox{blue!10}{%
    \hspace{0.25em}%
    \strut#1%
    \hspace{0.25em}%
  }%
  \hspace{-0.25em}}

\newcommand*{\codeltln}[1]{%
  \fboxsep0pt%
  \colorbox{blue!10}{%
    \strut#1%
  }}

\newcommand*{\figref}[1]{Fig.~\ref{fig:#1}}
\newcommand*{\secref}[1]{Sec.~\ref{sec:#1}}
\newcommand*{\apxref}[1]{Appendix~\ref{apx:#1}}

\newcommand*{\locref}[1]{Line~\ref{loc:#1}}
\newcommand*{\secarrow}[1]{$\rightarrow$ Sec.~\ref{sec:#1}}

\newcommand{\eg}{{{e.g.\@}}}

\newcommand{\ie}{{{i.e.\@}}}

\newcommand{\rust}{{{\textsc{Rust}\@}}}
\newcommand{\tla}{{{\textsc{TLA$^{+}$}\@}}}

\newcommand{\coq}{{{\textsc{Coq}\@}}}
\newcommand{\vst}{{{\textsc{VST}\@}}}
\newcommand{\deepspec}{{{\textsc{DeepSpec}\@}}}
\newcommand{\ironfleet}{{{\textsc{IronFleet}\@}}}
\newcommand{\ironsync}{{{\textsc{ironsync}\@}}}
\newcommand{\armada}{{{\textsc{Armada}\@}}}
\newcommand{\igloo}{{{\textsc{Igloo}\@}}}
\newcommand{\trillium}{{{\textsc{Trillium}\@}}}
\newcommand{\iris}{{{\textsc{Iris}\@}}}
\newcommand{\ltl}{{{\textsc{LTL}\@}}}
\newcommand{\verifythis}{{{\textsc{VerifyThis}\@}}}

\newcommand{\memcached}{{{\textsc{Memcached}\@}}}

\newcommand{\prusti}{{{\textsc{Prusti}\@}}}
\newcommand{\viper}{{{\textsc{Viper}\@}}}

\newcommand{\paxos}{{{\textsc{Paxos}\@}}}
\newcommand{\civl}{{{\textsc{CIVL}\@}}}
\newcommand{\dafny}{{{\textsc{Dafny}\@}}}

\newcommand{\declareRule}[2]{%
\expandafter\newcommand\csname#1#2Name\endcsname{\textsc{#1.#2}}%
\expandafter\newcommand\csname#1#2\endcsname{\hyperlink{rule:#1#2}{\textsc{#1.#2}}}%
\expandafter\newcommand\csname#1#2Def\endcsname[2]{\hypertarget{rule:#1#2}{\inference[\textsc{#1.#2}]{##1}{##2}}}}

\declareRule{L}{Discharge}
\declareRule{L}{Str}
\declareRule{L}{GlAcq}
\declareRule{L}{GlRel}
\declareRule{L}{Split}
\declareRule{L}{QSplit}
\declareRule{L}{QEmpty}

\begin{document}

\title{Refinement Proofs in Rust Using Ghost Locks}

\author{Aurel B\'il\'y}
  \orcid{0000-0002-9284-9161}
  \affiliation{%
    \institution{ETH Zurich}%
    \city{Zurich}%
    \country{Switzerland}}
  \email{aurel.bily@inf.ethz.ch}
\author{Jo\~ao C. Pereira}
  \orcid{0000-0003-4671-4132}
  \affiliation{%
    \institution{ETH Zurich}%
    \city{Zurich}%
    \country{Switzerland}}
  \email{joao.pereira@inf.ethz.ch}
\author{Jan Sch\"ar}
  \affiliation{%
    \institution{ETH Zurich}%
    \city{Zurich}%
    \country{Switzerland}}
\author{Peter M\"uller}
  \orcid{0000-0001-7001-2566}
  \affiliation{%
    \institution{ETH Zurich}%
    \city{Zurich}%
    \country{Switzerland}}
  \email{peter.mueller@inf.ethz.ch}

\begin{abstract}
  Refinement transforms an abstract system model into a concrete, executable program, such that properties established for the abstract model carry over to the concrete implementation. Refinement has been used successfully in the development of substantial verified systems. Nevertheless, existing refinement techniques have limitations that impede their practical usefulness. Some techniques generate executable code automatically, which generally leads to implementations with sub-optimal performance. Others employ bottom-up program verification to reason about efficient implementations, but impose strict requirements on the structure of the code, the structure of the refinement proofs, as well as the employed verification logic and tools.

  In this paper, we present a novel refinement technique that removes these limitations. It supports a wide range of program structures, data representations, and proof structures. Our approach supports reasoning about both safety and liveness properties. We implement our approach in a state-of-the-art verifier for the \rust{} language, which itself offers a strong foundation for memory safety. We demonstrate the practicality of our approach on a number of substantial case studies.
\end{abstract}

\maketitle

\section{Introduction}\label{sec:intro}
  Refinement is a technique to connect an abstract model of a system to another, lower-level formulation of the same system. It is especially valuable when specifying and verifying complex systems, such as distributed systems or multithreaded programs, because it allows one to prove invariants of the system as a whole on the level of the abstract model, then show that these invariants are preserved when the system is refined down its implementation-level components. Several recent developments of verified systems make use of refinement~\cite{CompCert_2006,sel4_2009,IronFleet_2015,DeepSpec_2017,Armada_2020}.

We focus on \emph{bottom-up refinement}, which connects a concrete implementation, written in a high-level programming language, to an abstract, mathematical model. With such an approach, the implementation can make use of features of modern programming languages, thus allowing for performant, maintainable code. By contrast, other refinement approaches either do not address the connection between the lowest-level abstract model and the implementation; or else use code extraction to generate executable code, which may lead to sub-optimal code.

In this paper, we present a novel methodology to prove that an implementation refines an abstract model given as a transition system. We defer a more thorough discussion of related work to \secref{relwork}, but to the best of our knowledge, our methodology is the first that enables \emph{flexible} refinement proofs along \emph{all} of the following dimensions:

\begin{itemize}
  \item \emph{Program structure}. We do not impose any particular
implementation structure. Refinement reasoning is localised to code locations which correspond to steps in the abstract model.
  A refinement proof can thus be added to an existing implementation, without rewriting the implementation itself.
  We address both multi-threaded processes and distributed systems.
  \item \emph{Automation and Scalability}. Our approach uses off-the-shelf, SMT-based deductive verifiers to discharge proof obligations. Verification of safety properties requires only modest annotation overhead. To ensure scalability, our approach is method- and thread-modular.
  \item \emph{Liveness properties}. Our approach addresses verification of liveness properties in a deductive verification setting. Proofs of liveness properties are constructed in ghost code integrated with the implementation and the safety part of the proof.
  \item \emph{Integration and Reusability}. Our approach is implemented in a state-of-the-art general-purpose deductive verifier. The specifications we define for standard library methods are reusable and integrate with the existing functional safety specifications.
\end{itemize}

\paragraph{Approach}

Our methodology embeds the abstract model to be refined into the implementation as ghost state, in the form of a transition system. The implementation then interacts with this model using a \emph{ghost lock}, which allows the program to \emph{acquire} a view of the abstract state and to later \emph{release} it, indicating which transition took place if the state was modified. Every release must then correspond to a legal transition. In other words, the ghost lock provides access to a two-state system invariant.

Each thread in a multi-threaded process, resp. each thread of every node of a distributed system, is given its own handle to the ghost lock. The ghost lock is erasable at compile time, thus it provides no synchronisation. As a result, operations within ghost lock critical sections must be linearisable.

To modularly reason about individual nodes of a system, we use \emph{guards}, which describe how each component may affect the global system state. As in existing guarded transition system reasoning~\cite{CAP_2010,Caper_2017}, this allows nodes to preserve local knowledge of the system across environment steps.

To prove liveness properties, we deeply embed \ltl~\cite{LTL_1977} formulas in the assertion language. These are then connected to obligations~\cite{DeadlockFreeChannels_2010,FiniteBlocking_2015}, abstract linear resources which must eventually be discharged by the holder, to prevent non-termination or lack of progress.

To integrate our methodology with real-world codebases, we instantiated it in \rust~\cite{Rust_2014}, a programming language which provides strong guarantees about memory safety with its ownership type system. It is a systems programming language and allows interaction with low-level data types and operations to maximise performance. These factors make it well suited to implement distributed systems, which often have high correctness, performance, and scalability requirements. The ownership systems allow direct extraction of separation-logic-style assertions from well-typed programs~\cite{Prusti_2019}. This reduces annotation overhead for proofs, facilitating further reasoning about functional correctness and making \rust{} an ideal target for deductive verification. The ownership system also facilitates refinement reasoning: for example, mutable references are known to not alias with other variables currently accessible by other threads, which means local data updates can be freely performed within ghost lock critical sections.

We evaluated our approach by implementing and verifying a simplified version of the \memcached{} caching system~\cite{Memcached_2004}, as well as several smaller case studies.
Our case studies and tool implementation will be submitted as an artefact.
The verification of \memcached{} has been posed as a challenge in the 2nd \verifythis{} long-term challenge presented at ETAPS 2023~\cite{VerifyThisLTC_2023}; we will submit our verified implementation there.

\paragraph{Contributions}

We make the following technical contributions:

\begin{itemize}
  \item We present a verification methodology for refinement proofs that offers more flexibility than prior work in terms of program structure, automation, and real-world integration (\secarrow{msafe}).
  \item We present a novel reasoning technique for local, control-flow sensitive guarded invariants (\secarrow{msafe:guards}).
  \item We present a novel connection of \ltl{} formulas to obligations, enabling the proof of both safety and liveness properties to be integrated into the executable code (\secarrow{mlive}).
  \item We implement our methodology in a state-of-the-art \rust{} verifier, resulting in strong automation, and the ability to reuse general-purpose verification techniques for parts of the proof unrelated to refinement (\secarrow{impl}).
  \item We evaluate our methodology on a number of case studies, demonstrating its expressiveness and ability to adapt to an evolving codebase (\secarrow{eval}).
\end{itemize}

\section{Methodology: Safety}\label{sec:msafe}
  In this section, we show how to prove safety properties using our approach on a running example. The example consists of two nodes interacting over a network. Node A maintains a counter and sends consecutive numbers to node B. In turn, node B performs some (presumably expensive) computation on the received number, then sends a message back, containing both the original request and the computed response.
A simplified implementation of both nodes in \rust{} is shown in \figref{impl}. Here, we assume that the communication channel has already been set up, and that the nodes will keep communicating indefinitely.
Due to the unpredictable nature of communication over a network, our implementation may exhibit different behaviours, including the following:
\begin{itemize}
  \item The messages from node A to B may be successfully delivered, and the responses may be computed and successfully delivered to A before it times out.
  \item The communication channels may lose messages. Node A will repeatedly send the same number until node B responds.
  \item Node A may time out and repeat its request
  before node B computes the response.
  Node B does not keep track of which requests it has already replied to, and may process the same message multiple times.
  In turn, node A may receive the same message multiple times. In that case, node A ignores stale messages.
\end{itemize}
Regardless of which communication behaviours are observed during the execution of the system,
the messages received (and responded to) by node B will always have a value less than or equal to node A's counter. This is
an invariant of the system and we show how to prove it in \secref{msafe:invariants}.

\begin{figure}
\begin{minipage}[t]{0.45\linewidth}
  \begin{rustcode}
// assume connection to node B in socket
fn node_a(socket: &mut TcpStream) {
  let mut ctr = 0;
  loop {
    // send
    (*\codehl{socket.send(ctr)}*)socket.send(ctr);(*\label{loc:impl:a-send}*)
    // wait for response, with timeout
    match (*\codehl{socket.recv\_timeout()}*)socket.recv_timeout()(*\label{loc:impl:a-recv}*) {
      Some((n, resp)) if ctr == n =>
        { ctr += 1; }
      None => {}
    }
  }
}\end{rustcode}
\end{minipage} \;
\begin{minipage}[t]{0.45\linewidth}
  \begin{rustcode}[15]
// assume connection to node A in socket
fn node_b(socket: &mut TcpStream) {
  loop {
    // receive (block)
    let num = (*\codehl{socket.recv()}*)socket.recv();(*\label{loc:impl:b-recv}*)
    // compute
    let resp = do_something(num);
    // respond
    (*\codehl{socket.send((num, resp))}*)socket.send((num, resp));(*\label{loc:impl:b-send}*)
  }
}\end{rustcode}
\end{minipage}
\Description{Code listings in Rust for node A and node B.}
\caption{Implementation of node A (left) and node B (right). The {\fboxsep0pt\colorbox{red!10}{\hspace{0.25em}\strut{}highlighted expressions\hspace{0.25em}}} are I/O operations discussed later in the text. For simplicity, we omit the (de)serialisation of transmitted data.}
\label{fig:impl}
\end{figure}

\subsection{Model definition}

In our methodology, we specify programs with transition systems directly in \rust{}.
Because of this, we are able to prove refinement in a general-purpose \rust{} verifier, without additional formalisms or tooling\footnote{To define the behaviour of a system abstractly, it is common to use a formalism such as \tla{}, where the model is defined as a transition system. For reference, we also provide a \tla{} specification for the example from \figref{impl} in \apxref{tla}. A mapping from our representation to \tla{} is discussed in \apxref{mapping}.}.
In general, a transition system consists of a set of states, a set of action labels, a set of initial states, and
a relation that describes the valid transitions from any state, given an action label. In this section,
we demonstrate how to define these components for our runnning example in \rust{}.

First, we define the set of states of the abstract model as a \code{struct} \code{SystemState} containing the current value of node A's counter (field \code{a\_ctr}), the message currently being processed by node B (\code{b\_work}), as well as the state of the channels in either direction (\code{a\_to\_b} and \code{b\_to\_a}).
Next, we define the action labels as an \code{enum}  \code{Action} containing a variant per kind of I/O operation performed by the system.
Variant \code{BSend} has an integer argument, which represents the response chosen by the implementation of node B\@.
The model does not constrain this value, which means it is existentially quantified in the specification; the value in \code{BSend} can be seen as the existential witness. The behaviour of the other actions is fully defined by the system state, as we will see shortly. The \code{ALoss} and \code{BLoss} actions model the behaviour of the \emph{environment}, accounting for message losses.

\begin{center}
  \begin{minipage}[t]{0.45\linewidth}\vspace{-10pt}
    \begin{rustcode}
struct SystemState {
  a_ctr: i32,
  b_work: Option<i32>,
  a_to_b: Seq<i32>,
  b_to_a: Seq<(i32, i32)>,
}
\end{rustcode}
  \end{minipage} \; \begin{minipage}[t]{0.45\linewidth}\vspace{-10pt}
    \begin{rustcode}
enum Action {
  ASend,      ARecv,
  BSend(i32), BRecv,
  ALoss,      BLoss,
}
\end{rustcode}
  \end{minipage}
\end{center}

\noindent
The usual \code{init} and \code{next} predicates, here as \rust{} functions, define valid initial states and transitions:

\begin{rustcode}
fn init(state: SystemState) -> bool {
  state.a_ctr == 0 (*\codepoint{A}*)
  && state.b_work == None (*\codepoint{B}*)
  && state.a_to_b == Seq::empty() && state.b_to_a == Seq::empty() (*\codepoint{C}*)
}
\end{rustcode}

\noindent The \code{init} predicate requires \codepoint{A} that the counter is initialised to zero, \codepoint{B} that node B is initially not performing any computation, and \codepoint{C} that both channels are empty.

\begin{rustcode}
fn next(p: SystemState, s: SystemState, a: Action) -> bool {
  match a {
    Action::ASend => s == SystemState {
      a_to_b: p.a_to_b.append(p.a_ctr), (*\codepoint{D}*)
      ..p }, // functional update syntax, uses p for all other fields
    Action::ARecv => p.b_to_a.len() > 0 (*\codepoint{E}*)
      && s == SystemState {
        b_to_a: p.b_to_a.tail(),
        a_ctr: p.a_ctr.max(p.b_to_a.head().0 + 1),
        ..p },
    Action::ALoss => p.b_to_a.len() > 0
      && s == SystemState {
        b_to_a: p.b_to_a.tail(),
        ..p },
    // ...
  }
}
\end{rustcode}

\noindent The \code{next} predicate\footnote{The full listing is provided in \apxref{full}.} is parameterised by the previous state \code{p}, the next state \code{s}, and the action \code{a} taken. The action \code{ASend} updates the network state by \codepoint{D} adding the current counter value to the outgoing channel. This action is always enabled, thus there are no constraints on \code{p}, unlike \code{ARecv}, which \codepoint{E} requires at least one message in the incoming channel.

All definitions provided in this section are embedded in the implementation as \emph{ghost code}, \ie{}, code that is added for specification purposes
and which does not interfere with regular code. Thus, it can be safely erased without observable differences in the program outcome.

\subsection{Updating the model state}
\label{sec:msafe:nonenv}

The model state includes the abstract state of each node and that of the environment. The implementation interacts with the environment by calling
methods that perform I/O, and which are provided by (unverified) libraries such as the \rust{} standard library.
In \figref{impl}, there are calls to three such methods: \code{send} in node A (\locref{impl:a-send}) and in node B (\locref{impl:b-send}), \code{recv_timeout} in node A (\locref{impl:a-recv}), and \code{recv} in node B (\locref{impl:b-recv}).
To describe the effect of these calls on the model state, we attach specifications, in the form of \emph{pre-} and \emph{postconditions} to
the corresponding methods\footnote{Following the \prusti{} syntax, the attributes \texttt{\#[requires(..)]} and \texttt{\#[ensures(..)]} denote pre- and postconditions,
respectively. The operator \texttt{old(..)} refers to the value of the given variable in the initial state of the method call, and \texttt{result} refers to the value returned by the call.}, as shown below for the methods \code{send} and \code{recv}:
\begin{center}
  \begin{minipage}[t]{0.45\linewidth}\vspace{-10pt}
    \begin{rustcode}
#[ensures(c == old(c).append(v))]
fn send<T>(v: T, c: &mut Seq<T>);\end{rustcode}
  \end{minipage} \; \begin{minipage}[t]{0.45\linewidth}\vspace{-10pt}
    \begin{rustcode}
#[ensures(!old(c).is_empty())]
#[ensures(c == old(c).tail()
  && result == old(c).head())]
fn recv<T>(c: &mut Seq<T>) -> T;\end{rustcode}
  \end{minipage}
\end{center}

\noindent In both methods, we expect a sequence denoting the state of the channel to be passed as an additional, ghost argument. This argument is used in
the postcondition of both methods to describe the effects of the method on the channel: \code{send} appends the message \code{c} to the sequence
denoting the channel, whereas \code{recv} removes the first element of the sequence and returns it.
Since it is a mutable reference, the verifier considers the value it points to to be modified to an arbitrary value that satisfies the postcondition.
Note that these specifications do not require the correct channel to be passed, \eg{}, node A may call \code{send} and mistakingly pass the state of the
buffer from node B to A; in \secref{msafe:gl}, we show an improved specification where the correspondence is checked by the verifier.

The responsibility to update the non-environment fields of the abstract model lies with the implementation. For example, node A must update \code{a_ctr} according to the transitions it performs. The second part of the loop in node A therefore performs a ghost update:

\begin{rustcode}
let state = /* ... */; // mutable reference to the system state
match socket.recv_timeout(&mut state.b_to_a) {
  Some((n, resp)) if ctr == n => {
    ctr += 1;
    (*\codehl{state.a\_ctr += 1;}*)state.a_ctr += 1; // ghost update
  }
  None => {}
}
\end{rustcode}

\noindent At this point, node A could assert that it is indeed performing the expected \code{ARecv} transition by referring to a copy of the state before the action and after the action:

\begin{rustcode}
assert!(next(prev_state, state, Action::ARecv));
\end{rustcode}

\subsection{Ghost lock}\label{sec:msafe:gl}
The model state abstracts over the global state of the system, including the environment. When we verify the
implementation of a component of the system, we must maintain, in ghost code, a view of the global state as an instance
of the model state; after all, methods that perform I/O may have their preconditions and postconditions defined in terms of the
model state.

To guarantee \emph{soundness}, \ie{},
that if the verifier succeeds, the implementation does indeed refine the abstract model, we must guarantee that
the view of the system maintained by the implementation is always consistent with the state of all other nodes and with the
state of the environment. In our methodology, we achieve this by \emph{sharing a single instance of the model state
among all nodes}.
Akin to locks, which control access to shared resources in multi-threaded programs, \emph{ghost locks} control access to the shared model state. Their interface is similar to regular locks, as shown below:

\begin{rustcode}
impl GhostLock {
  fn acquire(&mut self);
  fn release(&mut self, action: ActionKind);
  fn release_stutter(&mut self);
  fn locked(&self) -> bool;
  fn state(&mut self) -> &mut SystemState;
}
\end{rustcode}

Like a regular lock, a ghost lock may be \emph{acquired} to gain exclusive access to the (single instance of the) model state, and later \emph{released}, giving up the exclusive access to the model state. Unlike a regular lock, a ghost lock can only be used in \emph{ghost code}, thus calling its methods must not change the behaviour of the original program. As a consequence, ghost locks cannot be used for synchronisation.
Moreover, we treat each critical section of a ghost lock as an \emph{atomic} update of the model state~(or a stuttering step); thus, no intermediate state of the system should be observable. To ensure that this reasoning is sound, we require
all critical sections to be \emph{linearisable}~\cite{Linearisability_1990}. In our implementation, we reason about linearisability of statements
within critical sections using Lipton's reduction~\cite{LiptonReduction_1975}. Finally, we impose that, at any moment in time, each node running in the system has at most a single instance of the ghost lock. This ensures that no node can maintain two incompatible views of the system, and use them to derive contradictions.

The \code{acquire} operation marks the beginning of a critical section.
The \code{locked} method indicates whether the node is currently executing
in a critical section, and, if so, the method \code{state} provides a mutable reference to the system state through which updates can be performed.
The end of a critical section may be marked with a call to \code{release}.
All calls to \code{release} are annotated with the action performed in the critical section.
Firstly, this allows the programmer to communicate intent, \ie{}, that a particular action was intended and none other. Secondly, it makes it easier for automated verifiers to prove that the action took place, without relying on instantiating existential quantifiers, which would be needed for the \code{BSend} action, for example. Critical sections may also end with a call to \code{release_stutter}, to indicate that no transition took place. An explicit stutter release is useful as it avoids the need for a dedicated ``stutter'' action label. Since critical sections consist of paired calls rather than syntactic blocks, it is possible, for example, to conditionally release the ghost lock with different actions, depending on the results of operations within the critical section.

When a ghost lock is released, the node loses access to the shared state. The node may later
re-acquire the lock to access the model state again; in the time between releasing and re-acquiring
the lock, another node may have acquired it and modified the state of the system.
As such, when a node acquires the ghost lock, the model state is \emph{havocked},
\ie{}, the verifier assumes an arbitrary value for the abstract state. In \secref{msafe:guards}, we
show how to use invariants of the abstract model and \emph{guards} to soundly preserve knowledge about
the model state between different critical sections.

Having introduced ghost locks, we can now rewrite the specifications of the standard library methods shown in \secref{msafe:nonenv}
such that the user no longer has to pass (maybe erroneously) the components of the model state in each call. Instead,
these methods now receive a mutable reference to the \emph{acquired} ghost lock as a ghost argument. The environment can be accessed
through such a reference, allowing specifications on external methods to perform updates on the environment state. The specifications for \code{send} and \code{recv} which use the ghost lock could thus look as follows\footnote{These specifications are only suitable for node A; in our implementation there is a further mapping layer that allows different channels in the system to be treated uniformly without \emph{ad-hoc} specifications for each node or channel. We do not describe the layer in this paper, but the high-level idea is that the abstract model of a program is \emph{projected} to other transition systems related to particular features, \eg{}, the I/O channels; these are then linked to the main model based on a shared abstraction of the feature, \eg{}, sequences for I/O channels.}:

\begin{center}
  \begin{minipage}[t]{0.42\linewidth}\vspace{-10pt}
    \begin{rustcode}
#[requires(gl.locked())]
#[ensures(gl.locked())]
#[ensures(gl.state().a_to_b
  == old(gl.state().a_to_b)
    .append(v)))]
fn send(v: i32, gl: &mut GhostLock);\end{rustcode}
  \end{minipage} \; \begin{minipage}[t]{0.47\linewidth}\vspace{-10pt}
    \begin{rustcode}
#[requires(gl.locked())]
#[ensures(gl.locked())]
#[ensures(gl.state().b_to_a
    == old(gl.state().b_to_a).tail()
  && result
    == old(gl.state().b_to_a).head())]
fn recv(gl: &mut GhostLock) -> (i32, i32);\end{rustcode}
  \end{minipage}
\end{center}

With specifications provided for I/O methods, the verifier checks that any I/O operation is indeed performed within a ghost lock critical section.
The I/O operations receive the ghost lock as argument; their specifications  define how the environment state, accessible via the acquired ghost lock, was updated. This requires that the I/O operations must occur within ghost lock critical sections, that the critical sections are annotated with the correct action label, and that the implementation also updates the non-environment ghost state accordingly. The entire refinement proof thus naturally develops from the proof obligations resulting from using I/O methods, since any implementation must use I/O methods to interact with any other node or the environment.

A first attempt to update the running example with ghost lock annotations may be as follows (showing only the loop bodies, and assuming \code{gl} is a mutable reference to the ghost lock):

\begin{center}
\begin{minipage}[t]{0.45\linewidth}\vspace{-10pt}
  \begin{rustcode}
// send
gl.acquire();
socket.send(ctr, gl);
gl.release(Action::ASend);(*\codepoint{C}*)

// wait for response, with timeout
gl.acquire();
match socket.recv_timeout() {
  Some((n, resp)) => {
    if ctr == n {
      ctr += 1;
      gl.state().a_ctr += 1;
    }
    gl.release(Action::ARecv);(*\codepoint{D}*)
  }
  None => { gl.release_stutter();(*\codepoint{A}*) }
}
\end{rustcode}
\end{minipage} \;
\begin{minipage}[t]{0.45\linewidth}\vspace{-10pt}
  \begin{rustcode}
// receive
gl.acquire();
let num = socket.recv(gl);
gl.state().b_work = Some(num);
gl.release(Action::BRecv);(*\codepoint{E}*)

// compute
let resp = do_something(num);

// respond
gl.acquire();
socket.send((num, resp), gl);
gl.state().b_work = None;
gl.release(Action::BSend(resp));(*\codepoint{B}*)
\end{rustcode}
\end{minipage}
\end{center}

\noindent The environment state accessible through the acquired ghost lock is updated according to the specifications of I/O methods, by passing the ghost lock as a ghost argument. Non-environment ghost state updates are performed by the implementation; our methodology ensures that all updates to the model state
are linearisable, and that, when the lock is released,
the system state was correctly updated according to the label passed to the \code{release} method.

When node A does not receive a response from node B within some pre-determined timeout \codepoint{A}, the ghost lock is released with a \emph{stutter step} to indicate that no change occurred in the system state.  When releasing \codepoint{B} the ghost lock for the \code{BSend} action, the sent value is provided as an argument.

\subsection{Preserving local knowledge using guards}\label{sec:msafe:guards}

In the previous subsection, we show how to use ghost locks to ensure that all nodes have a consistent view of model state. Moreover, we annotate the running example with a ghost lock, from which a node may obtain permission to update the model state. Trying to verify the running example as is fails though, as the verifier is not able to prove that each critical section ending on a call to \code{release} corresponds to a valid step of the system. A closer look reveals the following problems:
\begin{itemize}
  \item \codepoint{C} For the \code{ASend} action to be valid, the sent value (\code{ctr}) must be equal to the value of the counter in the abstract state (\code{a_ctr}). However, the verifier conservatively assumes that there are other nodes which may modify \code{a_ctr}.
  \item \codepoint{D} The same problem occurs for the \code{ARecv} action, but there is a further complication: the update of \code{ctr} and \code{a_ctr} only follows the specification if node B is responding to a number less than or equal to the current \code{ctr} value.
  \item \codepoint{B}\codepoint{E} Node B cannot safely assume that node A does not update the field \code{b_work}.
\end{itemize}

\noindent At the root of these problems is the fact that, up to this point, we treated all nodes in the system as equal. This need not be true; in fact, complex systems are often made up of heterogeneous nodes, each of which may play a different role in the system, thus it can be expected that they perform different kinds of actions. Crucially, not every node may perform every transition.

To account for the heterogeneous nature of distributed systems, we use \emph{guards}, a common solution to reason about shared-state concurrency \cite{CAP_2010,Caper_2017}. Guards are affine resources owned by nodes, \ie{}, every guard has exactly one owner (an executing node or the environment), or it has no owner (the guard cannot be used anymore). Guards represent the permission to perform certain actions in the system. Ownership of a guard thus places an upper bound on what the environment might do, such that we can preserve information about the system state that is stable under environment interference.

In our example, we define three kinds of guards, for the two nodes and the environment, using a \rust{} \code{enum}\footnote{As with actions, the variants of this \code{enum} may contain data. Therefore, there can be unboundedly many guards in a system. This is useful, for example, in server applications, where each client-handling thread owns one guard instance.} and a function which determines whether a guard is needed for an action:

\begin{center}
\begin{minipage}[t]{0.21\linewidth}\vspace{-10pt}
  \begin{rustcode}
enum GuardKind {
  NodeA,
  NodeB,
  Environment,
}
\end{rustcode}
\end{minipage} \;
\begin{minipage}[t]{0.7\linewidth}\vspace{-10pt}
  \begin{rustcode}
fn guard_needed(a: Action, g: GuardKind) -> bool {
  match a {
    Action::ASend | Action::ARecv => g == GuardKind::NodeA,
    Action::BSend(_) | Action::BRecv => g == GuardKind::NodeB,
    Action::ALoss | Action::BLoss => g == GuardKind::Environment,
  }
}
\end{rustcode}
\end{minipage}
\end{center}

\noindent This definition states that guard \code{NodeA} is required for the \code{ASend} and \code{ARecv} actions, for example. Although in this system, there are only actions which require a single guard, in our case studies we also found situations where multiple guards are required to perform a single action. In such cases, the \code{guard_needed} function would contain a disjunction of the needed guard kinds for the action. The advantage over a single guard is that multiple guards can be (temporarily) owned by different nodes, as long as it is not necessary to perform actions which require their combination.

All guards of the system are created when the model is initialised and stored in a \emph{guard dispenser}. The verifier checks that each guard is only dispensed once. However, the affine nature of guards is already naturally represented and proven by the \rust{} type system: each guard is an owned value and using a guard is considered a mutating operation. Since safe \rust{} guarantees that there is only one mutable reference to any given value, only one node is able to use any guard at a time. The main operations of guards accessible within \rust{} are as follows:

\begin{rustcode}
impl Guard {
  fn kind(&self) -> GuardKind;
  fn open<F: Fn(SystemState) -> bool>(&mut self, gl: &GhostLock, pred: F);
  fn last_state(&self) -> SystemState;
}
\end{rustcode}

\noindent \code{kind} returns the kind of the guard, as defined earlier. Within a ghost lock critical section, the method \code{open} can be used to \emph{open} the guard for the duration of that critical section, \ie{}, to show that the executing node indeed mutably owns the guard at this point. A reference to the ghost lock is passed at the same time, allowing the verifier to check that the ghost lock is currently acquired. Any opened guard is closed when the critical section ends. The \code{open} method allows the node to perform actions protected by this guard.

Often, when opening a guard, it is useful to learn additional facts about the abstract state that can be justified by owning that very guard. In other words, there are invariants that hold in the system as long as transitions protected by the given guard (or set of guards) are \emph{not} taken.
As an example, suppose node A owns the guard \code{NodeA}. Given the definition of \code{guard_needed}, this means node B cannot perform the \code{ARecv} action. Consequently, if node A knows that the value of \code{a_ctr} is, for example, zero at the time it releases the ghost lock, then it must remain zero in the state when it subsequently re-acquires the ghost lock, because only the \code{ARecv} action modifies the value of \code{a_ctr}.
Such a claim is expressed in ghost code via the second argument of \code{open}. In the loop of node A, the local variable \code{ctr} and the ghost variable \code{a_ctr} should remain in sync. By the same line of reasoning as outlined above, the predicate expressing this equality is accepted by the verifier\footnote{In this paper we overload \rust{} closure syntax for predicates. Such closures cannot cause side-effects (otherwise, it is rejected by the type-checker), but may capture copies of the local state. To map such syntax to logical assertions, the captured variables are seen as the free variables of a formula, instantiated with some values at the point the \texttt{open\_with\_predicate} call takes place.}:

\begin{rustcode}
// send
gl.acquire();
guard.open(gl, |state| state.a_ctr == ctr); // ctr matches a_ctr
socket.send(ctr, gl); // this operation thus sends the correct value (i.e. a_ctr) to node B
gl.release(Action::ASend);
\end{rustcode}

The verifier checks, by induction, that the given predicate holds when the guard is opened:

\begin{enumerate}
  \item \emph{Base case}: the state in which the guard was last closed must satisfy the predicate. To this end, a copy of the state is recorded for each guard and accessed using the \code{last_state} function.
  \item \emph{Inductive case}: given two consecutive states related by an action which does \emph{not} require the guard, if the predicate holds in the first state, then it must hold in the second state.
\end{enumerate}

The reasoning principle is similar to VCC \emph{claims} \cite{VCCClaims_2010}, although unlike VCC claims, the guard does not fix a particular invariant -- the user may open the guard with arbitrary invariants. In our approach we record the \emph{state} in which the guard was last closed and defer checking the invariant until it is actually needed by the implementation. As a result, the implementation is free to choose a different predicate every time the guard is opened, where the predicate may make use of node-local data, or differ depending on the control flow taken.

In our methodology, guards may change owners. For example, a guard may be placed inside a lock. For this to be useful, the lock invariant must restrict the value of \code{last_state} in some way; otherwise, opening the guard with any non-trivial predicate would fail.

The \code{Environment} guard encompasses the permissions of the environment to perform im\-ple\-men\-ta\-tion-unrelated actions. In this example, this includes losing messages on either of the two communication channels (the \code{ALoss} and \code{BLoss} actions). This guard must not be given to any node (we forbid the construction of such a guard). Although defining an environment guard explicitly makes the model clearer, individual node implementations need not make any special assumptions due to its existence: it is sound to simply assume that any guard that the current node does \emph{not} hold may be used by another part of the system.

\subsection{Model invariants}\label{sec:msafe:invariants}

In this section, we explain how invariants of the model may be made available to the implementation.
As stated in \secref{msafe}, the messages received (and responded to) by node B will always have a value less than or equal to node A's counter. In the absence of integer overflow, this property seems intuitive: node B cannot respond to a request it has not received yet.
Although this property is not explicitly stated in the abstract model, it follows from the specification. If our specification was obtained from a \tla{} module, such a property could in principle instance be proved (for instance, using the \tla{} \emph{Proof System}~\cite{TLAPS_2010}) and simply assumed by our implementation. However, we instead justify the property within the verifier itself, which has the benefit that no additional tool or formalism is required. The proof can be decomposed into the following steps, tracing the flow of values from node A to node B and then back again:

\begin{enumerate}
  \item \code{a_ctr} is monotonically increasing.
  \item Numbers in \code{a_to_b} are at most \code{a_ctr}.
  \item If \code{b_work} is not \code{None}, the value it wraps is at most \code{a_ctr}.
  \item Numbers in the first component of \code{b_to_a} are at most \code{a_ctr}.
  \item The first component of all entries in \code{b_to_a} is at most \code{a_ctr}.
\end{enumerate}

Step (1) follows from the definition of \code{next}. Steps (2)--(4) are justified inductively: the property holds in the initial state and is preserved by any valid transition. Step (5) is a concrete application of step (4) for the callsite in node A. Within the verifier, each step is expressed as a lemma method \cite{VeriFast_2010} whose specification expresses the property of interest, and whose body provides the proof for that lemma. Using the theorem then corresponds to calling the lemma method.

\section{Methodology: Liveness}\label{sec:mlive}
 Up to this point, we have focused on the safety properties of the system: properties which can be proven to be preserved by each individual step of the system, \ie{}, when releasing the ghost lock. Verifying \emph{liveness properties}, on the other hand, poses multiple challenges. First, liveness properties concern infinite traces, so it is generally not possible to show that the property holds when performing a single ghost lock step. Second, in the case of reactivity properties\footnote{Using terminology from~\citet{TemporalPicture_1991}, reactivity properties are conjunctions of formulas of the form \( \always \eventually p \lor \eventually \always q \), when expressed in \ltl{}.}, there may never be a point at which the property is definitely satisfied; instead, it may be possible only to show that progress towards the property was made. Finally, it may not be possible to check the property locally in one node.

In our example, some liveness properties of interest include:

\begin{enumerate}
  \item Node A keeps sending requests to node B indefinitely.
  \item Node B will eventually respond to any request from node A.
  \item Every number will eventually be requested and responded to.
\end{enumerate}

In the following sections,
we describe how \ltl{} formulas are used to represent liveness properties in our methodology (\secarrow{mlive:formulas});
we introduce a mechanism to verify that the properties are \emph{eventually} satisfied (\secarrow{mlive:obl});
we discuss how the verifier is guided to prove liveness properties of a system (\secarrow{mlive:rules}); and
we discuss how fairness assumptions, such as ones about network behaviour, can be embedded into the proof (\secarrow{mlive:assumptions}).

\subsection{LTL formulas}\label{sec:mlive:formulas}

A common formalism to express liveness properties is linear temporal logic (\ltl{})~\cite{LTL_1977}. It is a logic where temporal properties are expressed using first-order logic combined with temporal operators, such as \emph{always} ($\always$) or \emph{eventually} ($\eventually$). In the model of abstract transition systems, \emph{always} is a universal quantification over all the subsequent states reached in the current computation. \emph{Eventually} is the existential counterpart. In this paper, we use the following fragment of \ltl{}:
\begin{align*}
  \phi &::= \phi \land \phi
    \;|\; \phi \lor \phi
    \;|\; \phi \rightarrow \phi &&\text{standard logical connectives} \\
  &\phantom{::=}|\; \always \phi
    \;|\; \eventually \phi
    \;|\; \phi \Rightarrow \phi &&\text{temporal operators} \\
  &\phantom{::=}|\; \bot
    \;|\; \top
    \;|\; S
    \;|\; A &&\text{atoms} \\
  &\phantom{::=}|\; \forall v.\, \phi
    \;|\; \exists v.\, \phi &&\text{value quantifiers}
\end{align*}

In the above, $S$ represents an arbitrary \emph{state formula}, a predicate over a single state of the abstract model. $A$ represents an arbitrary \emph{action formula}, a two-state predicate over consecutive states of the abstract model. We will use action labels, such as \code{ARecv}, as action formulas. The \emph{temporal entailment}, \( \phi_1 \Rightarrow \phi_2 \) represents \( \always ( \phi_1 \rightarrow \phi_2 ) \). In addition to the temporal operators (which quantify over states) we also allow quantification over \emph{values} (not connected to any particular state), which can bind free variables appearing in state formulas and action formulas.

The three liveness properties of our system can thus be represented as:

\begin{enumerate}
  \item \( \always \eventually \code{ASend} \)
  \item \( \always \eventually ( \exists r.\, \code{BSend(r)} ) \)
  \item \( \forall i.\, \eventually ( \exists r.\, \code{b_to_a' == b_to_a.append((i, r))} \land \code{BSend(r)} ) \)
\end{enumerate}

\noindent \code{ASend} and \code{BSend(r)} are action formulas corresponding to actions of the model. \code{b_to_a' == b_to_a} \code{.append((i, r))} is also an action formula, where \code{b_to_a'} refers to state of \code{b_to_a} \emph{after} the step. The formula has the free variables \code{i} and \code{r}, each of which is bound by one of the quantifiers.

In our implementation, we use a deep embedding of \ltl{} to specify liveness properties. Although we defer a more detailed discussion of this embedding until \secref{impl:ltl}, we will now describe how \ltl{} reasoning relates to the rest of our approach. Throughout the rest of this section, we will use a \codeltl{blue background} to denote \ltl{} formulas, to avoid the syntactic overhead of the deep embedding.

\subsection{Obligations}\label{sec:mlive:obl}

Obligations~\cite{DeadlockFreeChannels_2010,FiniteBlocking_2015,IronObligations_2019,ObligationsSync_2019}, are resources that must be explicitly discharged. Failure to discharge them, either by terminating the execution without calling an appropriate method to discharge the obligation or by never terminating are rejected by the proof system. In the previous subsection, we showed how we represent the liveness properties to be verified as \ltl{} formulas. In this subsection, we discuss how such formulas can be associated with obligations to guarantee that these properties are satisfied.

Every obligation must eventually be discharged by the node that holds it, \ie{}, there are finitely many steps before the obligation is fulfilled. To this end, each obligation is associated with a termination measure, a standard technique to turn liveness (termination in particular~\cite{Termination_1993}) into safety properties, amenable to automated verification. The measure is a function mapping obligations to a value in a well-founded set. Any method call or loop iteration entered with an obligation held must decrease the termination measure value for that obligation. Because the set is well-founded, there is no way to indefinitely delay discharging the obligation.

Specifications, such as method preconditions or loop invariants, may refer to obligations. Such an assertion is a resource assertion in separation logic. For example, an obligation in a method precondition indicates that a call to the method consumes the obligation. A call to the method from a scope where the obligation is not held violates the precondition.

An example of obligations in our methodology is a ghost lock release obligation. This obligation appears in the postcondition of the ghost lock acquire method, thus indicating that the node acquiring the ghost lock must release it. Symmetrically, the obligation also appears in the precondition of the ghost lock release method, but \emph{not} in its postcondition, indicating that the obligation is discharged by a call.

\begin{center}
\begin{minipage}[t]{0.35\linewidth}\vspace{-10pt}
  \begin{rustcode}
impl GhostLock {
  #[ensures(gl_release())]
  fn acquire(&mut self);
}
\end{rustcode}
\end{minipage} \;
\begin{minipage}[t]{0.55\linewidth}\vspace{-10pt}
  \begin{rustcode}
impl GhostLock {
  #[requires(gl_release())]
  fn release(&mut self, action: ActionKind);
}
\end{rustcode}
\end{minipage}
\end{center}

\subsection{LTL proof rules}\label{sec:mlive:rules}

\begin{figure}

\begin{center}
\scriptsize\begin{tabular}{cc}
  \LDischargeDef{%
    \codeltl{\( \phi(\code{i}) \)}%
  }{%
    \{ \texttt{show\_at(}\codeltln{\( \phi \)}\texttt{, i)} \}%
    \; \texttt{discharge()} \;%
    \{ \}}\vspace{0.2cm}
  & \LStrDef{%
    \codeltl{\( \phi_1(\texttt{i}) \)} \Rightarrow \codeltl{\( \phi_2(\texttt{j}) \)}%
  }{%
    \{ \texttt{show\_at(}\codeltln{\( \phi_2 \)}\texttt{, j)} \}%
    \; \texttt{strengthen()} \;%
    \{ \texttt{show\_at(}\codeltln{\( \phi_1 \)}\texttt{, i)} \}}\vspace{0.2cm}
\end{tabular}

\begin{tabular}{c}
  \LSplitDef{%
  }{%
    \{ \texttt{show\_at(}\codeltln{\( \phi_1 \land \phi_2 \)}\texttt{, i)} \}%
    \; \texttt{split()} \;%
    \{%
      \texttt{show\_at(}\codeltln{\( \phi_1 \)}\texttt{, i)}%
      * \texttt{show\_at(}\codeltln{\( \phi_2 \)}\texttt{, i)} \}}\vspace{0.4cm} \\
  \LQSplitDef{%
    P(v)%
  }{%
    \{ \texttt{show\_at(}\codeltln{\( \forall i.\, P(i) \Rightarrow A(i) \)}\texttt{, i)} \; \}%
    \; \texttt{qsplit()} \;%
    \{%
      \texttt{show\_at(}\codeltln{\( A(v) \)}\texttt{, i)}%
      * \texttt{show\_at(}\codeltln{\( \forall i.\, i \neq v \land P(i) \Rightarrow A(i) \)}\texttt{, i)} \}}\vspace{0.4cm} \\
  \LQEmptyDef{%
    \forall i.\, \lnot P(i)
  }{%
    \{ \texttt{show\_at(}\codeltln{\( \forall i.\, P(i) \Rightarrow A(i) \)}\texttt{, i)} \; \}%
    \; \texttt{qempty()} \;%
    \{ \}}\vspace{0.2cm}
\end{tabular}\end{center}
\Description{Rules of the LTL proof system related to obligations and univeral quantifier reasoning. L.Discharge discharges an LTL obligation if the formula holds. L.Str strengthens an LTL formula to be shown. L.Split splits an obligation to show a conjunction of two formulas into two obligations to show the conjuncts. L.QSplit splits off a conjunct off a quantified formula. Note that we impose an additional restriction on the order of conjuncts in the quantifier, not shown here. L.QEmpty discharges a quantifier which holds vacuously.}
\caption{Rules of the \ltl{} proof system related to obligations and universal quantifier reasoning. \protect\LDischarge{} discharges an \ltl{} obligation if the formula holds. \protect\LStr{} strengthens an \ltl{} formula to be shown. \protect\LSplit{} splits an obligation to show a conjunction of two formulas into two obligations to show the conjuncts. \protect\LQSplit{} splits off a conjunct off a quantified formula. Note that we impose an additional restriction on the order of conjuncts in the quantifier, not shown here. \protect\LQEmpty{} discharges a quantifier which holds vacuously.}
\label{fig:ltl-rules}
\end{figure}

To prove that the implementation satisfies a given liveness property, we associate the \ltl{} formula representing that property with an obligation given to the implementation. The obligation is discharged when it can be shown that the associated \ltl{} formula holds. The truth value of an \ltl{} formula depends on the current state and the subsequent states in the trace.
As an example, the action formula \code{ASend} is true for consecutive states of the trace, related by a \code{ASend} transition.

With a global view of the trace, states can be indexed with natural numbers. \codeltl{\( \phi ( n ) \)} then refers to the truth value of the \ltl{} formula \( \phi \) in the \( n \)-th state of the trace. Although individual nodes do not have a global view of the trace (\ie{}, the intermediate states visited during an environment step are not known), successive ghost lock steps within the same node always correspond to states further in the trace. Consequently, we omit the \emph{next} operator in our \ltl{} fragment, as it is not useful when reasoning about a single node in a distributed system: any environment step performs zero or more transitions in the general case, invalidating any local knowledge about precise step counts. We will refer to the obligation to show \codeltl{\( \phi ( \code{i} ) \)}, \ie{}, that \( \phi \) holds at trace index \code{i} as \code{show_at(}\codeltln{\( \phi \)}\code{, i)}.

The atoms of \ltl{} formulas over abstract states specify the state changes performed in critical sections of the ghost lock, which are the only changes of the abstract state. The method \code{id()}, accessible on an acquired ghost lock, returns a natural number corresponding to the current index in the trace. We define state formulas as predicates over the abstract state at the point of release, and action formulas as two-state predicates over the states at the points of acquire and release.

We provide the rules of an \ltl{} proof system as a library of ghost methods. A selection of these rules is shown in \figref{ltl-rules}. The implementation can then use calls to ghost methods to manipulate \ltl{} obligations, thus proving the \ltl{} formulas. As an example, we consider the first property, \codeltl{\( \always \eventually \code{ASend} \)}, and show it holds in the implementation of node A:

\begin{rustcode}
let mut show_from = 0;(*\codepoint{A}*)
loop {
  invariant!(show_at((*\codeltln{\( \always \eventually \)ASend}*), show_from));
  gl.acquire();
  let id = gl.id(); // id >= show_from (*\codepoint{B}*)
  // obligations: show_at((*\codeltln{\( \always \eventually \)ASend}*), show_from)
  strengthen();(*\codepoint{C}*)
  // obligations: show_at((*\codeltln{\( \eventually \)ASend}*), show_from) * show_at((*\codeltln{\( \always \eventually \)ASend}*), show_from + 1)
  strengthen();(*\codepoint{D}*)
  // obligations: show_at((*\codeltln{ASend}*), id) * show_at((*\codeltln{\( \always \eventually \)ASend}*), show_from + 1)
  strengthen();(*\codepoint{E}*)
  // obligations: show_at((*\codeltln{ASend}*), id) * show_at((*\codeltln{\( \always \eventually \)ASend}*), id + 1)
  guard.open(gl, |state| state.a_ctr == ctr);
  socket.send(ctr, gl);
  gl.release(Action::ASend); // thus (*\codeltl{ASend(id)}*) is true
  discharge();(*\codepoint{F}*)
  // obligations: show_at((*\codeltln{\( \always \eventually \)ASend}*), id + 1)
  show_from = id;(*\codepoint{G}*)
  // obligations: show_at((*\codeltln{\( \always \eventually \)ASend}*), show_from + 1)
  // (remainder of the loop implementation)
}
\end{rustcode}

In the above, we \codepoint{A} introduce the ghost variable \code{show_from} to keep track of progress in the \ltl{} property, and then use \code{show_from} in the obligation in the loop invariant. \codepoint{B} This variable is a lower bound on the index of any ghost lock critical section appearing in this loop, \ie{}, \code{id} is greater or equal to \code{show_from}.

We use three calls to the \code{strengthen} ghost method to rewrite the obligations into more concrete, stronger goals\footnote{Once again, to avoid the syntactic complexity of the deep embedding, we omit the arguments to these calls and only present the obligations in comments.}. First, \codepoint{C} we split the \emph{always} operator into its first state (at index \code{show_from}) and the rest of the states. Second, \codepoint{D}, we concretise the \emph{eventually} operator, since we will perform the \code{ASend} action in the current ghost lock step. Finally, \codepoint{E}, we push the second obligation to a later state to fit the shape of the loop invariant at the end of the loop.

After the ghost lock step, \codepoint{F} the action formula is known to hold at this index, thus the first obligation can be discharged. \codepoint{G} After updating the \code{show_from} variable, the remaining obligation is consumed by the loop invariant, and the proof is completed.

Quantifiers with an infinite domain are useful to express properties of non-terminating programs. The third property of our system becomes the loop invariant of node B:

\begin{rustcode}
let mut sent = 0;(*\codepoint{H}*)
loop {
  invariant!(show_at((*\codeltln{\( \forall \texttt{i}.\, \)i >= sent\( \; \Rightarrow \eventually ( \exists \texttt{r}.\, \)b\_to\_a' == b\_to\_a.append((i, r))\( \; \land \; \)BSend(r)\( ) \)}*), 0));
  // receive
  gl.acquire();
  guard.open(gl, |state| state.b_work == None);
  let num = socket.recv(gl);
  gl.state.b_work = Some(num);
  gl.release(Action::BRecv);
  // compute
  let resp = do_something(num);
  // respond
  gl.acquire();
  let id = gl.id();
  guard.open(gl, |state| state.b_work == Some(num));
  socket.send((num, resp), gl);
  gl.state.b_work = None;
  gl.release(Action::BSend(resp));
  // obligations: show_at((*\codeltln{\( \forall \texttt{i}.\, \)i >= sent\( \; \Rightarrow \eventually ( \exists \texttt{r}.\, \)b\_to\_a' == b\_to\_a.append((i, r))\( \; \land \; \)BSend(r)\( ) \)}*), 0)
  if num > sent {
    qsplit();    (*\codepoint{I}*)
    // obligations: show_at((*\codeltln{\( \eventually ( \exists \texttt{r}.\, \)b\_to\_a' == b\_to\_a.append((num, r))\( \; \land \; \)BSend(r)\( ) \)}*), 0)
    //   * show_at((*\codeltln{\( \forall \texttt{i}.\, \)i >= sent + 1\( \; \Rightarrow \eventually ( \exists \texttt{r}.\, \)b\_to\_a' == b\_to\_a.append((i, r))\( \; \land \; \)BSend(r)\( ) \)}*), 0)
    strengthen();(*\codepoint{J}*)
    // obligations: show_at((*\codeltln{b\_to\_a' == b\_to\_a.append((num, resp))\( \; \land \; \)BSend(resp)}*), id)
    //   * show_at((*\codeltln{\( \forall \texttt{i}.\, \)i >= sent + 1\( \; \Rightarrow \eventually ( \exists \texttt{r}.\, \)b\_to\_a' == b\_to\_a.append((i, r))\( \; \land \; \)BSend(r)\( ) \)}*), 0)
    discharge(); (*\codepoint{K}*)
    sent += 1;
    // obligations: show_at((*\codeltln{\( \forall \texttt{i}.\, \)i >= sent\( \; \Rightarrow \eventually ( \exists \texttt{r}.\, \)b\_to\_a' == b\_to\_a.append((i, r))\( \; \land \; \)BSend(r)\( ) \)}*), 0)
  } else { (*\codepoint{L}*) }
}
\end{rustcode}

In the above, we \codepoint{H} once again introduce a ghost variable, this time to track the domain of the quantifier in the \ltl{} formula. Once node B has received a number, computed the response, and sent it to node A, we check if the number has not been sent by node B before\footnote{The entire condition and obligation manipulation is ghost code, thus this state need not be stored by the actual implementation.}. If so, \codepoint{I} we split the first conjunct off the quantifier. \codepoint{J} We concretise the \emph{eventually} operator and pick a witness for the existential quantifier, allowing us to \codepoint{K} discharge the conjunct. After updating the ghost variable, the obligation matches the one in the loop invariant again.

In the case that the number has been seen by node B before (branch \codepoint{L}), we must still show that progress is made towards the \ltl{} property overall. We omit the details in this paper, but our proof system contains rules for discharging progress based on the guaranteed behaviour of other nodes.

The use of ghost methods to describe proof steps allows the full expressiveness of (our fragment of) \ltl{}, at the cost of some automation. As an example, the verifier will automatically infer that \codeltl{\( \phi_1 \land \phi_2 \)} holds when \codeltl{\( \phi_1 \)} and \codeltl{\( \phi_2 \)} both hold, but it will not find witnesses for existential quantifiers.

\subsection{Fairness assumptions}\label{sec:mlive:assumptions}

Some of the liveness properties of our system can be justified only if the network is not faulty. If the channel from A to B is faulty and always loses messages, node A may keep sending requests (and thus satisfy the first property), but the system overall does not progress (the third property). Similarly, if the channel from B to A always loses messages, node may B keep responding to requests (the second property), but once again, the system overall does not progress.

Although the network is part of the model, it is not part of our implementation, so we express the expected behaviour as assumptions. We prove liveness properties under the \emph{strong fairness assumption} that any message repeatedly sent to a channel will eventually be delivered, which we can represent as the formulas \codeltl{\( \eventually \always (\code{b_to_a.len() == 0}) \lor \always \eventually \code{ARecv} \)} and \codeltl{\( \eventually \always (\code{a_to_b.len() == 0}) \lor \always \eventually \code{BRecv} \)}. These formulas are assumed when calling network I/O methods, to justify later \ltl{} proof steps. In the example of node B above, the second formula, combined with the first property of the system (\codeltln{\( \always \eventually \code{ASend} \)}), justifies why the loop in node B is guaranteed to see every number at some point.

There are other, lower-level assumptions, for instance, that the 
underlying execution environment and thread scheduler work correctly.
These assumptions are built into our verification technique for concurrent programs and, thus, not explicit in the specifications.

\section{Implementation}\label{sec:impl}
  We implemented our approach in the state-of-the-art deductive \rust{} verifier, \prusti~\cite{Prusti_2019}, based on \viper~\cite{Viper_2016}, a framework for automated separation logic reasoning. To support our methodology, we had to extend \prusti{} with support for obligations. Our methodology is not inherently tied to \prusti{}-specific reasoning, thus it should be implementable in other \rust{} verifiers.

\subsection{Deep embedding of LTL}\label{sec:impl:ltl}

In this section we briefly describe our embedding of \ltl{} into \rust{} with \prusti{} annotations. \ltl{} formulas are encoded as types which implement the \code{Ltl} trait. This trait is parameterised by the abstract model (described in \secref{impl:model}), and has one associated function\footnote{The \prusti{} annotation \texttt{\#[pure]} indicates that a function is deterministic, side-effect-free, and terminating. Pure functions can be used within specifications of other functions and ghost code.}:
\begin{rustcode}
trait Ltl<M: Model> {
  #[pure] fn holds(&self, gl_id: Nat) -> bool;
}
\end{rustcode}

\noindent The \code{holds} function defines whether the current \ltl{} formula holds in the given state, which is identified by the trace index of the ghost lock critical section (discussed in \secref{mlive:rules}).

The obligations are then generic over types of this trait:
\begin{rustcode}
obligation! { fn show_at<M: Model, L: Ltl<M>>(l: L, id: Nat); }
\end{rustcode}

Discharging the obligation is accomplished by calling a method which consumes the obligation, but only if the precondition \( \phi\code{.holds(gl_id)} \) is satisfied.

Atoms in \ltl{} consist of terminal formulas, such as \code{True} or \code{False}, which are implementations of the \code{Ltl} trait with a trivial \code{holds} implementation; one-state predicates on the abstract model's state; and two-state predicates on the state, indicating that some transition has happened. The user can invoke a function-like macro to declare a type which represents a state formula, which internally implements the \code{Ltl} trait with a suitable specification on the \code{holds} function.

Composite \ltl{} formulas are types which combine other \ltl{} formulas. Such combinators are generic over the sub-formulas, as can be seen in the definition of conjunction:
\begin{rustcode}
struct Conj<M: Model, L1: Ltl<M>, L2: Ltl<M>>(L1, L2);
\end{rustcode}

\noindent As a result, the conjunction \( \phi_1 \land \phi_2 \) is represented by a different type than the conjunction \( \phi_2 \land \phi_1 \).

The library also provides a number of methods to manipulate \ltl{} formulas. In particular, a rewriting system allows exchanging the obligation to show an \ltl{} formula for the obligation to show a stronger \ltl{} formula, which corresponds to the \LStr{} rule. Because the shape of \ltl{} formulas is fully expressed in the type, such rewriting is often based purely on types, though sometimes additional preconditions are added.

\subsection{Model trait}\label{sec:impl:model}

The mapping of our methodology to a \rust{} library makes heavy use of the trait system. The \code{Model} trait defines an abstract model:
\begin{rustcode}
trait Model {
  type AbsState: Copy;     (*\codepoint{A}*) type Action: Copy;  (*\codepoint{B}*) type GuardKind: Copy;(*\codepoint{C}*)
  type Liveness: Ltl<Self>;(*\codepoint{D}*) type EnvState: Copy;(*\codepoint{E}*)
  #[pure] fn get_env_state(s: Self::AbsState) -> Self::EnvState; (*\codepoint{F}*)
  #[pure] fn init(s: Self::AbsState) -> bool;
  #[pure] fn next(s: Self::AbsState, n: Self::AbsState, a: Self::Action) -> bool;
  #[pure] fn guard_needed(a: Self::Action, g: Self::GuardKind) -> bool;
  #[pure] fn init_ltl(s: Self::AbsState) -> Self::Liveness;
}
fn new<M: Model>() -> (GhostLock<M>, GuardDispenser<M>, Ltl<M>) { .. } (*\codepoint{G}*)
\end{rustcode}

\noindent Any implementation of this trait corresponds to an abstract model. The implementation declares \codepoint{A} the type of full system states, \codepoint{B} the type of labelled transitions, \codepoint{C} the type of guard kinds, and \codepoint{D} the shape of the \ltl{} formula representing the liveness property.

The associated type \code{EnvState} \codepoint{E} and the required method \code{get_env_state} \codepoint{F} together define a subset of the full system state that represents the state of the environment. This part of the state can only be modified by calling trusted methods, such as methods of the standard library. The remainder of the required methods maps naturally to methods already explained in \secref{msafe}.

Finally, the provided \code{new} method \codepoint{G} initialises an instance of the ghost lock, a \emph{guard dispenser} with all the guards of the system, and the liveness property. The lock can be used as described before, while the dispenser can be used to obtain specific guards.

\section{Evaluation}\label{sec:eval}
  \begin{figure}
  \begin{tabular}{@{}lllcrrrr@{}} \toprule
    Sec. & \multicolumn{2}{c}{Program} & Liveness & \( \#_C \) & \( \#_M \) & \( \#_P \) & VT (s) \\\midrule
    \ref{sec:eval:mc} & \memcached{}~\cite{Memcached_2004} & v1 & \checkmark & 117 & 225 & 286 & 334.7 \\
    & & v2 & \checkmark & 118 & 225 & 290 & 354.6 \\
    & & v3 & \checkmark & 181 & 235 & 377 & 379.7 \\
    & & TCB & & 320 & -- & 57 & -- \\
    \ref{sec:eval:queue} & Prod./cons. queue~\cite{VerusTransitionSystems_2022} & \textsc{SeqCst} & \checkmark & 125 & 169 & 247 & 392.8 \\
    & & \textsc{AcqRel} & & 125 & 169 & 428 & 654.5 \\
    \ref{sec:eval:other} & \paxos{}~\cite{Paxos_1998} & & & 105 & 111 & 205 & 217.7 \\
    \ref{sec:eval:other} & Lock-free hash set~\cite{FPSet_2017} & & & 54 & 134 & 361 & 280.6 \\\bottomrule
  \end{tabular}
  \Description{Table of evaluation results with 8 lines.%
4 lines relate to the memcached case study, including three different versions and the trusted code base. The memcached versions have between 117 and 181 lines of code, between 225 and 235 lines of model, between 286 and 377 lines of specifications/proofs, and verify in 334.7 to 379.7 seconds. The trusted code base is 320 lines of code and 57 lines of specifications.%
2 lines relate to the producer/consumer queue case study, one for sequentially consistent atomics, one for acquire-release atomics. The two versions both have 125 lines of code and 169 lines of model, and have 247 lines of specifications/proofs and 428 lines of specifications/proofs, respectively, and verify in 392.8 and 654.5 seconds, respectively.%
1 line relates to the Paxos case study, with 105 lines of code, 111 lines of model, 205 lines of specifications/proofs, verifying in 217.7 seconds.%
1 line relates to the Lock-free hashset case study, with 54 lines of code, 134 lines of model, 361 lines of specifications/proofs, verifying in 280.6 seconds.}
  \caption{Evaluation results. \( \#_C \) is the number of lines of code in the implementation, \ie{}, non-ghost code that is required for the implementation to work. We exclude empty lines, comments, import statements. \( \#_M \) is number of lines in the model definition, including the \code{Model} trait implementation and any associated types. \( \#_P \) is the number of lines of specifications and ghost code operations. For lines with a checkmark, \( \#_P \) also includes the annotations required for the liveness proof. The verification time (\textbf{VT}) is the 10\% Winsorised mean of the wall-clock runtime across 10 verification runs using \prusti{} commit \texttt{79d48686}, measured on an Intel Core i9-10885H 2.40GHz CPU with 16 GiB of RAM.}
  \label{fig:eval}
\end{figure}

To evaluate our approach, we implemented a number of concurrent and distributed systems from the literature, then specified and verified them. The results of our evaluation are shown in~\figref{eval}.

Before we discuss each case study in detail, we summarise the overall findings. The general annotation overhead is in the usual range for SMT-based verifiers. Interestingly, the liveness specification and proof add only around 10\% to both annotation and verification time overhead in both the \memcached{} and queue programs, even though they require manual applications of proof rules.
Our models are larger than one might expect; however, around 40\% (in the case of \textsc{SeqCst} queue) of the model lines are boilerplate, much of which could be generated by a macro or otherwise reduced by further focusing on a better user-facing interface. Despite their size, models are easier to review because they are declarative and contain fewer details.
The verification times are comparable to those of similarly sized codebases in prior \prusti{} work. As noted in \secref{impl}, our approach is not tied to \prusti{} itself and could be instantiated in other verifiers, as long as they support obligation-based reasoning, which may lead to performance improvements.

\subsection{\texorpdfstring{\memcached{}}{Memcached}}\label{sec:eval:mc}

\newcommand{\cmdSet}{{{\textsc{Set}\@}}}
\newcommand{\cmdGet}{{{\textsc{Get}\@}}}
\newcommand{\cmdDelete}{{{\textsc{Delete}\@}}}
\newcommand{\cmdCAS}{{{\textsc{CompareAndSwap}\@}}}

\memcached{} is an in-memory key-value store which can be accessed through a network protocol.
At the most basic view, it stores a mapping from keys to values, where both keys and values are byte sequences.
The protocol offers \cmdSet{}, \cmdGet{}, and \cmdDelete{} commands.

We developed a simplified version of \memcached{} in \rust{}, and proved that it refines an abstract model using our approach.
It is executable and interoperable with the original \memcached{} for the subset of features that we implemented.
We developed it in multiple iterations, each adding features that are present in the original implementation.
This iterative process allowed us to demonstrate that our approach is suitable for refinement proofs in an evolving codebase.
The protocol parser and serialiser are trusted, since such code was not the focus of this work.

\paragraph{First version (v1)}

The first version uses \rust{}'s built-in \code{HashMap} data structure to store key-value pairs, and only supports the \cmdSet{}, \cmdGet{}, and \cmdDelete{} commands.
This version is not concurrent yet; it handles one connection at a time.
The state and action definitions of the model are shown here:

\begin{center}
\begin{minipage}[t]{0.45\linewidth}\vspace{-10pt}
\begin{rustcode}
enum ConState {
  Idle,
  HaveCommand(AbsCmd),
  HaveResponse(AbsRes),
}
struct AbsState {
  con_cmd: Map<ConId, Seq<AbsCmd>>,
  con_res: Map<ConId, Seq<AbsRes>>,
  con_state: Map<ConId, ConState>,
  cache: Map<Seq<u8>, Option<Seq<u8>>>,
}
\end{rustcode}
\end{minipage} \;
\begin{minipage}[t]{0.45\linewidth}\vspace{-10pt}
\begin{rustcode}
enum Action {
  SendCommand(ConId, AbsCmd),
  ReceiveCommand(ConId, AbsCmd),
  SendResponse(ConId, AbsRes),
  ReceiveResponse(ConId, AbsRes),
  ProcessCommand(ConId, AbsCmd, AbsRes),
}
enum GuardKind {
  Storage,
  Connection(ConId),
}
\end{rustcode}
\end{minipage}
\end{center}

\code{AbsCmd} and \code{AbsRes} are abstract representations of commands (requests) and responses, respectively.
Each incoming connection is identified by a \code{ConId}.
As in the running example in \secref{msafe}, we represent the incoming and outgoing channels as sequences.
Upon accepting a client, the implementation enters a loop, in which the actions \code{ReceiveCommand}, \code{ProcessCommand}, and \code{SendResponse} are performed in turn.
The current step for the client is tracked using the \code{enum ConState}.
The \code{SendCommand} and \code{ReceiveResponse} actions are performed by clients connecting to the server, thus we model them as environment actions.
Our case study crucially relies on guards to couple the current point in the receive-process-send loop of each client to the \code{ConState} in the model state, using that client's \code{Connection(ConId)} guard.
The \code{ProcessCommand} action is additionally protected by the \code{Storage} guard, which maintains the link between the concrete \code{HashMap} and the abstract \code{Map} in the \code{cache} field.
The verification of this first version was straightforward, and involved adding the ghost lock, guard annotations, and ghost updates, keeping track of state in pre- and postconditions and loop invariants, and maintaining the connection between abstract and concrete versions of data structures.
The predicates used when opening guards are trivial, since the respective parts of the model state are fully determined by local state.

The liveness property which we prove is \codeltl{\( \forall \code{con}.\, \always \eventually \exists \code{res}.\, \code{SendResponse(con, res)} \)}.
To be able to show this, we assume that there is always eventually a new client connection, and that each client always eventually sends a command.
These assumptions are encoded by assuming that the \code{accept} and \code{read} methods terminate; such assumptions can be lifted with a more precise model of the network.
We also ignore error handling in the liveness verification.

\paragraph{Concurrent connections (v2)}
In a later version, we spawn a thread for each incoming connection, which runs the receive-process-send loop.
The \code{HashMap} is protected by a lock, which is only acquired for the \code{ProcessCommand} action.
We also put the \code{Storage} guard into the lock, such that it stays together with the concrete state for which it maintains the coupling to the model state.
This coupling invariant is then part of the lock invariant.
The model definition is unchanged compared to the previous version without concurrency, which shows that our technique can treat concurrency as an implementation detail that is introduced during refinement, but not reflected in the model.

\paragraph{Fine-grained locking (v3)}
In \memcached{}, the storage is not protected by a single global lock; instead, each hash table bucket is protected by a separate lock.
To implement this, we can no longer use \code{HashMap}, and instead need our own hash table implementation.
Our hash table is a vector of buckets, each inside a lock.
Each bucket contains a linked list of items, which stores the key, value, and pointer to the next item.

We no longer have a single \code{Storage} guard, but one \code{StorageBucket(usize)} guard for each bucket.
This guard is needed for a \code{ProcessCommand} action if the key that the command operates on hashes to this bucket.
Consequently, the predicate used when opening a \code{StorageBucket} guard is not a simple equality anymore, but a quantifier:

\begin{rustcode}
bucket.guard.open(gl, |state|
  forall(|key: Seq<u8>| hash_abs(key) == hash ==>
    item_valid(bucket.list, key, state.cache.get(key))));
\end{rustcode}

Even though the structure of our implementation changed significantly, \ie{}, we added concurrency and fine-grained locking to the initially sequential implementation, the model remains unchanged, apart from the guard definitions. This demonstrates that our approach allows for great flexibility in program structure.

The difference in verification times for the successive versions is small. Due to the modular nature of the verifier and our methodology, methods which are not changed need not be re-verified, so with the use of caching\footnote{Note that~\figref{eval} shows the full verification times, without enabling the cache.}, interactive development and verification in such a codebase is possible.

\subsection{Producer/consumer queue}\label{sec:eval:queue}

Our next case study is a single-producer, single-consumer queue, taken from the documentation for Verus Transition Systems~\cite{VerusTransitionSystems_2022}.
This case study demonstrates that our refinement approach can support various concurrency primitives, including atomics with weak ordering guarantees, and that data structures built with these primitives can be verified. We also demonstrate (with the \code{VerifiedCell} component below) that our refinement methodology is suitable for ensuring safety in the presence of \rust{} \code{unsafe} code.

Unlike \memcached{}, this case study is centred around a single data structure.
The queue consists of a vector storing the content of the queue, and head and tail pointers.
It is accessed through the \code{Producer} and \code{Consumer} handles, which are created when the queue is constructed.
To allow the \code{Producer} and \code{Consumer} to be used from different threads, the head and tail pointers are stored in atomic variables, which allows them to be accessed concurrently.
In \rust{}, a data structure which is shared between threads allows only read access by default.
The queue elements, however, are written by one thread and later read by another.
Typically, this would require the use of a wrapper data structure which allows \emph{interior mutability}, such as a \code{Mutex<T>}, which would ensure synchronisation and mutual exclusion, and then allow mutable access to the contained \code{T}.
However, in the queue example, the atomic accesses to the head and tail pointers already provide sufficient synchronisation, so the additional synchronisation overhead of the mutex is unnecessary.
Instead, the unverified code uses \code{UnsafeCell<T>}, which provides raw interior mutability.
Any access to the contents of the \code{UnsafeCell} must be wrapped in an \code{unsafe \{ ... \}} block, which indicates that safety must be checked manually be the user.

To allow verification of code which uses \code{UnsafeCell}, we introduce \code{VerifiedCell<T>}, a wrapper around \code{UnsafeCell} which relies on verification for checking the safety requirements, and is thus safe to use.
Concretely, access to an \code{UnsafeCell} is only safe if there are no data races.
A program has a \emph{data race} if there are two accesses to the same location, where at least one is a write, at least one is non-atomic, and there exists no happens-before relation between the accesses~\cite{ISOcpp_2020}.
\emph{Happens-before} is a partial order of all memory accesses.

For each atomic operation (read or write call), and each non-atomic access to a \code{VerifiedCell}, we define an abstract operation identifier.
We then encode the happens-before relation as a binary predicate between these identifiers.
The \code{VerifiedCell} accessor method ensures memory safety by requiring that happens-before holds between subsequent accesses.

We verified two versions of the queue, one with sequentially consistent and one with acquire-release atomics.
Acquire-release atomics offer better performance, but are more difficult to reason about, as evidenced by the increased amount of annotations needed to verify this version of the queue.
The key difference compared to sequentially consistent atomics in terms of the abstract model is that we store a \emph{set} of operation identifiers for each atomic variable, representing all write operations performed so far, as opposed to only the single identifier of the last write operation.

\subsection{Other case studies}\label{sec:eval:other}

Finally, we also specified and verified a version of the distributed consensus algorithm \paxos{}~\cite{Paxos_1998}, as well as a lock-free \emph{hash-set}~\cite{FPSet_2017}, used in the implementation of the TLC model checker. In both cases we were able to verify an executable implementation with relatively low specification overhead and acceptable verification times.

\section{Related Work}\label{sec:relwork}
  Various approaches~\cite{Chapar_2016,Velisarios_2018,DistributedProtocols_2018,Raft_2016} develop implementations that are correct by construction by
refining abstract models within \coq{} and then extracting executable OCaml programs.
Similarly, \citet{Maude_2020} model distributed systems in Maude's rewriting logic and compile them into
implementations running in distributed Maude sessions. The code extracted by these approaches is typically sub-optimal
(for instance, does not use mutable data structures) and cannot interface with existing libraries,
which is often necessary in practice.
In contrast, our methodology uses bottom-up verification and can
handle efficient implementations using concurrency, distribution, and node-local mutable state.

\trillium{}~\cite{Trillium_2021} is a refinement technique based on separation logic. Like our methodology, it does not impose
strict requirements on the structure of the implementations, and the function of invariants in \trillium{} is similar to our ghost lock. Furthermore, because \trillium{} is based on \iris{} and formalised in \coq{},
it provides an expressive specification language and enables foundational correctness proofs. On the other hand,
proofs in this framework are not easily automatable and require extensive manual work. Instead, our methodology
represents abstract models in first-order logic and automates verification using an SMT solver.

\armada{}~\cite{Armada_2020} supports the verification of concurrent, high-performance code written in a C-like language. To achieve refinement against an abstract model,
the user specifies a sequence of steps to gradually transform the implementation into the specification. Non-trivial refinement steps require complex \dafny{}~\cite{Dafny_2010} proofs showing a connection between two state machines. Unlike \armada{}, our methodology does not convert programs to state machines and the coupling between the abstract model and the implementation can be much looser.
The \civl{} verifier~\cite{CIVL_2015,CIVL_2020} also organises the refinement proof into multiple layers. Each layer is a structured concurrent program, where
the concurrent behavior is reflected in the program structure. This structure simplifies the proof obligations and allows automation, but also reduces program flexibility. Refinement steps are based on a set of trusted tactics. By contrast, our methodology imposes no restrictions on the program or proof structure.
\igloo{}~\cite{Igloo_2020} connects abstract models to concrete implementations via dedicated I/O specifications~\cite{IO_2015}. Similarly to our work, they support a variety of separation logics to reason about concrete implementations. However, their technique has not been shown to allow for threads performing I/O operations concurrently, whereas we have shown that our methodology has no such limitation.

Similar to our methodology, \ironfleet{}~\cite{IronFleet_2015} embeds abstract models as ghost state into executable programs
and automates verification using an SMT-based verifier, in their case \dafny{}. However, their refinement technique imposes severe
restrictions on how programs are structured. In particular, the programs must be sequential and their structure must mirror
the structure of the abstract model. Like \ironfleet{}, our approach supports verifying liveness properties. Unlike our temporal library, their API to prove temporal properties is proven correct against a model of the system's behaviour. On the other hand, \ironfleet{} makes use of \emph{always-enabled} actions to guarantee certain liveness properties by construction. It is unclear how this reasoning can be extended to unbounded transitions or guard-based reasoning.

\ironsync{}~\cite{IronSync_2023} also embeds the abstract model as ghost state, and, like us, uses ownership to reason
about accesses to the model state. In their approach, the abstract model is decomposed
into \emph{shards}, \ie{}, resources that provide access to a partial view of the global state. Importantly, shards can only be \emph{owned}
by a single thread. Actions in the system are specified in terms of the parts of the state that they access. Thus, a thread may only
perform an action if it owns all the shards that the action depends on -- this is similar to how we use guards to reason
about allowed actions. Guards allow for a more precise accounting of the allowed actions:
two actions may modify the same part of the abstract state; in \ironsync{}, it is not clear how to express that
only one action may have taken place when a thread does not hold the shard. In our approach, we express this by keeping the
ownership of one of the guards, but not the other.
Finally, their work focuses only on safety properties, whereas we support both safety and liveness properties.
We believe that our novel technique of tying \ltl{} formulas to obligations could be applied to their setting to support liveness properties.

The refinement technique~\cite{ITrees_2019} used in \deepspec{}~\cite{DeepSpec_2017} is based on the Verified Software Toolchain (\vst{})~\cite{VST_2018}, a framework for verifying C programs via a separation logic embedded in \coq{}. Instead of transition systems, they specify the intended system behavior using interaction trees~\cite{ITrees_2020}, which are embedded into \vst{}'s separation logic. In contrast, our methodology allows us to apply standard separation logics and existing program verifiers.
\citet{MessagePassing_2019} embed process calculus models into a concurrent SL, which is automated using Viper. Their refinement approach preserves state assertions, but it is unclear whether arbitrary trace properties are preserved.

\section{Conclusion}\label{sec:con}
  In this paper, we have introduced a novel methodology for refinement proofs of programs written in a high-level language that refine an abstract transition system model. The methodology centres around the use of ghost locks to allow flexible program structure and concurrency, showing both safety and liveness proofs. We have implemented our approach in the \prusti{} verifier, and evaluated our approach on several case studies, including an implementation of \memcached{}, demonstrating that the approach is expressive, amenable to automation, and performant. As future work, we plan to more thoroughly address initialisation in distributed systems; to provide better automation for the \ltl{} fragment; and to formalise our approach.

\bibliographystyle{ACM-Reference-Format}
\bibliography{paper}

\appendix

\section{\tla{} specification of running example}\label{apx:tla}
  \begin{minipage}[l]{\linewidth}\small\tlatex
\@x{}\moduleLeftDash\@xx{ {\MODULE} Example}\moduleRightDash\@xx{}%
\@x{ {\EXTENDS} Naturals ,\, Sequences}%
\@x{ {\VARIABLES} ACtr ,\, BWork ,\, BWorkNum ,\, AToB ,\, BToA}%
\@x{}\midbar\@xx{}%
 \@x{ Vars \.{\defeq} {\langle} ACtr ,\, BWork ,\, BWorkNum ,\, AToB ,\, BToA {\rangle}}%
\@pvspace{8.0pt}%
\end{minipage} \\
\begin{minipage}[l]{\linewidth}
  \begin{minipage}[t]{0.45\linewidth}\small\tlatex
    \@x{ Init\@s{4.08} \.{\defeq} \.{\land} ACtr \.{=} 0}%
    \@x{\@s{35.81} \.{\land} BWork \.{=} {\FALSE}}%
    \@x{\@s{35.81} \.{\land} BWorkNum \.{=} 0}%
    \@x{\@s{35.81} \.{\land} AToB \.{=} {\langle} {\rangle}}%
    \@x{\@s{35.81} \.{\land} BToA\@s{0.25} \.{=} {\langle} {\rangle}}%
    \@pvspace{8.0pt}%
    \@x{ ASend \.{\defeq} \.{\land} AToB \.{'}}%
     \@x{\@s{50.00} \.{=} Append ( AToB ,\, ACtr )}%
     \@x{\@s{43.48} \.{\land} {\UNCHANGED} {\langle} ACtr ,\, BWork , }%
     \@x{\@s{105.00} BWorkNum ,\, BToA {\rangle}}%
    \@pvspace{8.0pt}%
    \@x{ BSend\@s{0.39} \.{\defeq} \.{\land} BWork \.{=} {\TRUE}}%
    \@x{\@s{43.48} \.{\land} BWork \.{'} \.{=} {\FALSE}}%
    \@x{\@s{43.48} \.{\land} \exists \, Resp \.{\in} Nat : BToA \.{'} }%
     \@x{\@s{50.00} \.{=} Append ( BToA ,\, {\langle} BWorkNum, Resp {\rangle} )}%
     \@x{\@s{43.48} \.{\land} {\UNCHANGED} {\langle} ACtr ,\, BWorkNum ,\,
     AToB {\rangle}}%
    \@x{}%
    \@pvspace{8.0pt}%
    \@x{ ALoss\@s{0.84} \.{\defeq} \.{\land} BToA \.{\neq} {\langle} {\rangle}}%
    \@x{\@s{43.48} \.{\land} BToA \.{'} \.{=} Tail ( BToA )}%
     \@x{\@s{43.48} \.{\land} {\UNCHANGED} {\langle} ACtr ,\, BWork , }%
     \@x{\@s{105.00} BWorkNum ,\, AToB {\rangle}}%
    \@pvspace{8.0pt}%
    \@x{ Next \.{\defeq} \.{\lor} ASend}%
    \@x{\@s{35.85} \.{\lor} ARecv}%
    \@x{\@s{35.85} \.{\lor} BSend}%
    \@x{\@s{35.85} \.{\lor} BRecv}%
    \@x{\@s{35.85} \.{\lor} ALoss}%
    \@x{\@s{35.85} \.{\lor} BLoss}%
    \@pvspace{8.0pt}%
  \end{minipage}\begin{minipage}[t]{0.45\linewidth}\small\tlatex
    \@x{ TypeInv \.{\defeq} \.{\land} ACtr \.{\in} Nat}%
    \@x{\@s{49.03} \.{\land} BWork \.{\in} {\BOOLEAN}}%
    \@x{\@s{49.03} \.{\land} BWorkNum \.{\in} Nat}%
    \@x{\@s{49.03} \.{\land} AToB \.{\in} Seq ( Nat )}%
    \@x{\@s{49.03} \.{\land} BToA\@s{0.25} \.{\in} Seq ( Nat \times Nat )}%
    \@pvspace{8.0pt}%
    \@x{ ARecv\@s{0.84} \.{\defeq} \.{\land} BToA \.{\neq} {\langle} {\rangle}}%
    \@x{\@s{43.48} \.{\land} ACtr \.{'} \.{=} Max ( Head ( BToA ) [1] + 1 ,\, ACtr )}%
    \@x{\@s{43.48} \.{\land} BToA \.{'} \.{=} Tail ( BToA )}%
     \@x{\@s{43.48} \.{\land} {\UNCHANGED} {\langle} BWork ,\,
     BWorkNum ,\, AToB {\rangle}}%
    \@pvspace{8.0pt}%
    \@x{ BRecv\@s{1.23} \.{\defeq} \.{\land} BWork \.{=} {\FALSE}}%
    \@x{\@s{43.48} \.{\land} AToB \.{\neq} {\langle} {\rangle}}%
    \@x{\@s{43.48} \.{\land} BWork \.{'} \.{=} {\TRUE}}%
    \@x{\@s{43.48} \.{\land} BWorkNum \.{'} \.{=} Head ( AToB )}%
    \@x{\@s{43.48} \.{\land} AToB \.{'} \.{=} Tail ( AToB )}%
     \@x{\@s{43.48} \.{\land} {\UNCHANGED} {\langle} ACtr ,\, BToA
     {\rangle}}%
    \@pvspace{8.0pt}%
    \@x{ BLoss\@s{0.84} \.{\defeq} \.{\land} AToB \.{\neq} {\langle} {\rangle}}%
    \@x{\@s{43.48} \.{\land} AToB \.{'} \.{=} Tail ( AToB )}%
     \@x{\@s{43.48} \.{\land} {\UNCHANGED} {\langle} ACtr ,\, BWork , }%
     \@x{\@s{105.00} BWorkNum ,\, BToA {\rangle}}%
    \@pvspace{8.0pt}%
    \@x{ Live\@s{1.65} \.{\defeq} \.{\land} {\SF}_{ Vars} ( ASend )}%
    \@x{\@s{35.85} \.{\land} {\WF}_{ Vars} ( BSend )}%
    \@x{}%
    \@x{ Spec\@s{1.46} \.{\defeq} Init \.{\land} {\Box} [ Next ]_{ Vars}
    \.{\land} Live}%
    \@pvspace{8.0pt}%
  \end{minipage}
\end{minipage} \\
\begin{minipage}[l]{\linewidth}\small\tlatex
\@x{}\midbar\@xx{}%
\@x{ {\THEOREM} Spec \.{\implies} {\Box} TypeInv}%
\@x{}\bottombar\@xx{}%
\end{minipage}

\section{Full code listing of the next predicate}\label{apx:full}
  \begin{rustcode}
fn next(p: SystemState, s: SystemState, a: Action) -> bool {
  match a {
    Action::ASend => s == SystemState {
      a_to_b: p.a_to_b.append(p.a_ctr),
      ..p
    },
    Action::ARecv => p.b_to_a.len() > 0
      && s == SystemState {
        b_to_a: p.b_to_a.tail(),
        a_ctr: p.a_ctr.max(p.b_to_a.head().0 + 1),
        ..p
      },
    Action::ALoss => p.b_to_a.len() > 0
      && s == SystemState {
        b_to_a: p.b_to_a.tail(),
        ..p
      },
    Action::BSend(resp) => match p.b_work {
      Some(req) => s == SystemState {
        b_to_a: p.b_to_a.append((req, resp)),
        b_work: None,
        ..p
      },
      None => false,
    },
    Action::BRecv => p.a_to_b.len() > 0
      && p.b_work == None
      && s == SystemState {
        a_to_b: p.a_to_b.tail(),
        b_work: Some(p.a_to_b.head()),
        ..p
      },
    Action::BLoss => p.a_to_b.len() > 0
      && s == SystemState {
        a_to_b: p.b_to_a.tail(),
        ..p
      },
  }
}
\end{rustcode}

\section{Relation to \tla{} and other formalisms}\label{apx:mapping}
  We give a brief overview of the correspondence between our transition systems and ones defined in formalisms like \tla{}.

\subsection*{State, predicates}

The \code{AbsState} type of the model trait (\code{SystemState} in the running example from \secref{msafe}) combines all the variables of the abstract system with ones related to the environment. In \tla{}, both categories map to \textsc{variables} declarations.

The \code{init} predicate of our system also naturally maps to the \emph{Init} predicate in \tla{}.

Within the \code{next} predicate, our approach prescribes a more specific form than \tla{}. In \tla{}, \emph{Next} can be an arbitrary two-state predicate, although it is common that it is a formula that is a disjunction of existentially quantified conjunctions (actions). In our approach, the outer disjunction is represented by the \code{enum} type of action labels. Furthermore, the existential quantifiers are replaced by data stored in the \code{enum} variants. As noted in \secref{msafe:gl}, this form of actions was chosen for better automation, and for the programmer to be able to better communicate intent when transitions take place.

\subsection*{Types}

Our model definitions are embedded directly in \rust{}, a strongly typed language. \tla{} is not a typed language, though it is common for \tla{} modules to contain a definition of a type invariant which must be preserved by any transition.

The types used in our case studies, such as natural numbers and abstract sequences, sets, or maps, all have a counterpart in \tla{}. It would thus be possible to obtain a definition of the type invariant from the types in our model.

\subsection*{Guards}

Unlike in our approach, guard-based reasoning is not a first-class feature in \tla{}. It is, of course, possible to encode guards as ghost values within a \tla{} module. However, the concrete semantics of such guard algebras can vary from module to module.

\subsection*{Liveness}

The \ltl{} formulas we use to express liveness properties can directly map to \tla{}, since it also uses \ltl{} for temporal reasoning.

\subsection*{Summary}

In summary, mapping from our approach to \tla{} (or similar transition system-based formalism) is possible and automatable. A mapping in the other direction would require the source model to have additional annotations, or else depend on heuristics, \eg{} to infer types for variables based on the definition of actions.

\end{document}